\documentclass[12pt]{article}
\usepackage{graphics,cite}
\def \lsim
{\raisebox{-3pt}{$\>\stackrel{<}{\scriptstyle\sim}\>$}}
\def \gsim
{\raisebox{-3pt}{$\>\stackrel{>}{\scriptstyle\sim}\>$}}

\newcommand{\msbar}{$\overline{\mbox{MS}}$ }

\begin{document}

\begin{titlepage}

\begin{flushright}
\end{flushright}

\vspace{1.3cm}

\begin{center}
{\Large \bf\boldmath
Bottom-quark fragmentation:\\
comparing results from tuned event generators\\
\vspace{0.2cm}
and resummed calculations}
\end{center}

\vspace{5mm}

\begin{center}
{\large \bf G.~Corcella$^1$ and V.~Drollinger$^2$}\\

\vspace{5mm}

{$^1${\sl Department of Physics, CERN}\\ 
{\sl Theory Division}\\
{\sl CH-1211 Geneva 23, Switzerland}\\}

\vspace{3mm}

{$^2${\sl Dipartimento di Fisica Galileo Galilei, Universit\`a di Padova}\\
{\sl and INFN, Sezione di Padova}\\
{\sl Via Marzolo 8, I-35131 Padova, Italy}}

\end{center}

\par \vspace{2mm}
\begin{center}
{\large \bf Abstract}
\end{center}
\begin{quote}
  \pretolerance 10000

We study bottom-quark fragmentation in $e^+e^-$ annihilation, top
and Higgs decay $H\to b\bar b$,
using Monte Carlo event generators,
as well as calculations, based on the formalism of
perturbative fragmentation functions, which resum soft- and collinear-radiation
effects  in the next-to-leading
logarithmic approximation. We consider the PYTHIA and HERWIG generators, and 
implement matrix-element corrections to the parton shower simulation of
the $H\to b\bar b$ process in HERWIG.
We tune the Kartvelishvili, string and cluster
models to $B$-hadron data from LEP and SLD,
and present results in both $x_B$ and moment spaces.
The $B$-hadron spectra yielded by 
HERWIG, PYTHIA and resummed calculations show small discrepancies,
which are due to the different
approaches and models employed and to the quality of the
fits to the $e^+e^-$ data.
\end{quote}
\end{titlepage}

\section{Introduction}
Heavy-flavour phenomenology is currently one of the most lively fields
of investigation, in both experimental and theoretical particle physics.
In this paper, we consider $B$-hadron
production in $e^+e^-$ annihilation ($e^+e^-\to b\bar b$), 
top decay ($t\to bW$) and
the decay of the Standard Model Higgs boson $H\to b\bar b$.
In fact, data on $b$-flavoured hadron production from
SLD \cite{sld} and LEP \cite{aleph,opal} experiments
have been available for a few years, and they can be used to predict
$b$-quark fragmentation in other processes.
Bottom fragmentation in top decay is indeed one of the main sources
of uncertainties on the top-mass reconstruction
at the Tevatron \cite{tevatron} and the LHC. In particular, the 
analysis in Ref.~\cite{avto}, 
where the top-quark mass is determined from the
invariant mass of three leptons (one lepton from the $W$ decay and two leptons
from a $J/\psi \rightarrow \ell^+ \ell^-$ decay), depends on a proper
description of the fragmentation of $b$ quarks into
$B$ hadrons which decay further to $J/\psi$'s.
As for $H\to b\bar b$, the detection of this process
is impossible in the channel 
$gg(q\bar q)\to H\to b\bar b$ at hadron colliders, because of the large
irreducible $b\bar b$ background. The solution to this problem is to look
for $H\to b\bar b$ in associated production channels. The most relevant
search channels are Higgs production in association with a vector
boson at the Tevatron \cite{tevhig}, and in association with a $W$ boson
\cite{denegri} or with $t\bar t$ pairs \cite{muller} at the LHC.
In such channels, the 
$H\to b\bar b$ decay mode can be probed for a Higgs mass
up to $m_H\lsim 135$~GeV.
For heavier Higgs bosons, the decay to $b\bar b$ might be accessible in
Two-Higgs Doublet Models, with enhanced Higgs couplings to bottom quarks
\cite{atltdr}. In all cases, a good understanding of bottom fragmentation 
in Higgs decays is important for the Higgs mass reconstruction and the 
identification of $b$-jets, which both depend on the properties of the 
Higgs decay products.

In order to perform accurate searches and measurements, the use of 
precise QCD calculations
will be fundamental. 
In particular, fixed-order calculations are reliable
enough to predict total cross sections or widths, but
differential distributions
exhibit terms, corresponding to collinear or soft parton radiation,
that need to be summed to all orders to obtain a meaningful result. 
An example is the large term $\sim\alpha_S\ln(Q^2/m^2)$, 
where $Q$ is the hard scale of the
process and $m$ the heavy-quark mass, which appears in the 
energy distribution of a heavy quark at next-to-leading order (NLO).

A possible approach to study $b$-quark production in the considered
processes is based on the perturbative fragmentation formalism
\cite{mele}.
The NLO heavy-quark energy spectrum is expressed as the 
convolution of a coefficient function, describing the emission of a 
massless parton, and a perturbative fragmentation function $D(m_b,\mu_F)$,
associated with the transition of a massless parton into a massive
quark. 
The dependence of $D(m_b,\mu_F)$ on the factorization scale $\mu_F$
is determined by using the Dokshitzer--Gribov--Lipatov--Altarelli--Parisi
(DGLAP) evolution equations \cite{ap,dgl}, once an initial condition at a scale
$\mu_{0F}$ is given. The initial condition
of the perturbative fragmentation function, first computed in
\cite{mele}, has been proved to be process-independent in \cite{cc}.
Solving the DGLAP evolution equations we can resum the large 
$\ln(Q^2/m^2)$ (collinear resummation).

Moreover, heavy-quark energy distributions present terms that are enhanced 
when the energy fraction $x$ approaches 1.
Such contributions can be resummed
following the general method of \cite{sterman,ct} (soft or threshold
resummation).
Soft and collinear resummations in the next-to-leading logarithmic 
approximation (NLL) have been implemented for $b$ production in
$e^+e^-$ annihilation \cite{mele,cc}, top-quark  \cite{cm,ccm}
and Higgs \cite{cor} decays, in the framework of perturbative fragmentation
functions.

An alternative approach to address heavy-quark production and resum terms 
corresponding to soft and colliner radiation consists in using
Monte Carlo event generators.
Standard Monte Carlo programs, such as HERWIG \cite{her} or 
PYTHIA \cite{pythia62,pythia63}, simulate multiple emissions in the soft or
collinear approximation, and are provided with matrix-element
corrections \cite{mike,miu} to describe hard or large-angle radiation.

In order to describe hadronic spectra, perturbative calculations and
Monte Carlo parton showers need to be supplemented by models describing the 
hadronization. Calculations based on the fragmentation-function approach
typically predict hadron-level energy distributions convoluting
the partonic spectra with a non-perturbative fragmentation function,
such as the Kartvelishvili \cite{kart}
or the Peterson \cite{pet} models. These models 
contain parameters that need to be fitted to experimental data.
More recently \cite{canas}, it was suggested that one can fit directly
the $N$-moments of heavy-hadron cross sections and extract the moments
of the non-perturbative fragmentation function, with no need to assume
any functional form in $x$-space.
Likewise, Monte Carlo event generators simulate the hadronization using
appropriate non-perturbative models, such as the cluster model \cite{cluster}
and the string model \cite{string}, for HERWIG and PYTHIA, respectively.

The outline of this paper is the following.
In section 2 we review collinear and soft resummations in the
framework of perturbative fragmentation functions.
In section 3 we discuss Monte Carlo parton shower algorithms.
Section 4 describes matrix-element corrections
to parton-shower simulation and, in particular, the implementation
of such corrections to $H\to b\bar b$ processes in HERWIG.
In sections 5 and 6 we shall present our results on $B$-hadron
production in $e^+e^-$ annihilation, top and Higgs decays, in
$x$ and moment space respectively, after tuning the
hadronization models to $e^+e^-$ data from LEP and SLD.
Section 7 summarizes the main results and gives some concluding remarks.

\section{Perturbative fragmentation functions}
In this section we shall review the main points of the
perturbative fragmentation approach and of the implementation of collinear
and soft resummations. 
We shall consider $b$ production at next-to-leading order in
$Z$, top or Higgs decays, i.e.
\begin{equation}
P(Q) \to b(p_b) \bar b(p_{\bar b}) \left( g(p_g)\right),
\end{equation}
with $P=Z$ or $H$, and
\begin{equation}
t(Q) \to b(p_b)  W(p_W) \left( g(p_g)\right).
\end{equation}
The energy spectrum of a massive $b$ quark is given 
by the following general result:
 \begin{equation}
{1\over{\Gamma_0}}{{d\Gamma}\over {dx_b}}=\delta(1-x_b)
+{{\alpha_S(\mu)}\over{2\pi}}
\left[P_{qq}(x_b)\ln{{Q^2}\over{m_b^2}}+A(x_b)\right] +
{\cal O} \left( {{m_b^2}\over{Q^2}}\right)^p  .
\label{massb}
\end{equation}
In Eq.~(\ref{massb}), $x_b$ is the normalized $b$-quark energy fraction:
\begin{equation}
x_b={1\over{1-w}}{{2p_b\cdot Q}\over{Q^2}},
\label{xb}
\end{equation}
where $w=0$ in $Z$ and $H$ decay,
and $w=m_W^2/m_t^2$ in top decay, since the $W$ mass is not 
negligible with respect to the top mass.
Furthermore, in Eq.~(\ref{massb}) $\Gamma_0$ is the width of the Born process,
$\mu$ is the renormalization scale, $p\geq 1$,
$P_{qq}(x_b)$ is the Altarelli--Parisi splitting function:
\begin{equation}
P_{qq}(x_b)=C_F\left( {{1+x_b^2}\over {1-x_b}}\right)_+.
\label{pqq}
\end{equation} 
Function $A(x_b)$ is process-dependent and does not contain the $b$ mass.

The large logarithm $\ln(Q^2/m_b^2)$, which appears in Eq.~(\ref{massb}),
can be resummed following the approach of perturbative fragmentation
functions \cite{mele}.
The $b$ spectrum is expressed as the convolution of a massless coefficient
function and a perturbative fragmentation function $D(m_b,\mu_F)$,
associated with the transition of a massless parton into a heavy quark:
\begin{eqnarray}
{1\over {\Gamma_0}} {{d\Gamma_b}\over{dx_b}} (x_b,Q,m_b) &=&
\sum_i\int_{x_b}^1
{{{dz}\over z}\left[{1\over{\Gamma_0}}
{{d\hat\Gamma_i}\over {dz}}(z,Q,\mu,\mu_F)
\right]^{\overline{\mathrm{MS}}}
D_i^{\overline{\mathrm{MS}}}\left({x_b\over z},\mu_F,m_b \right)} \nonumber \\
&+& {\cal O}\left((m_b/Q)^p\right) \; .
\label{pff}
\end{eqnarray}
In Eq.~(\ref{pff}), $d\hat\Gamma_i /dz$ is the differential width for the 
production of a massless parton $i$ in any of the considered processes,
after subtraction of the collinear singularity in the \msbar factorization
scheme. Throughout this paper, and for all the
considered processes, we shall neglect secondary 
$b$-quark production via $g\to b\bar b$ splitting, which means that,
in Eq.~(\ref{pff}),
$i=b$ and $D_b^{\overline{\mathrm{MS}}}$ expresses the fragmentation
of a massless $b$ into a massive $b$.
The NLO \msbar coefficient functions were calculated 
in \cite{mele,cm,cor} for $Z$, $t$ and $H$ decays respectively.

The perturbative fragmentation function follows the 
DGLAP evolution equations
\cite{ap,dgl}. Its value at a given scale $\mu_F$ can be obtained once
an initial condition is given. In \cite{mele} the initial condition
$D_b^{\rm ini}(x_b,\mu_{0F},m_b)$ was calculated and its process-independence 
was established on more general grounds in \cite{cc}.
It is given at NLO by:
\begin{equation}
D_b^{\rm ini}(x_b,\mu_{0F},m_b)=\delta(1-x_b)+
{{\alpha_S(\mu_0^2)C_F}\over{2\pi}}
\left[{{1+x_b^2}\over{1-x_b}}\left(\ln {{\mu_{0F}^2}\over{m_b^2}}-
2\ln (1-x_b)-1\right)\right]_+.
\label{dbb}
\end{equation}

As discussed in \cite{mele}, solving the DGLAP equations for an evolution
from $\mu_{0F}$ to $\mu_F$, with a NLO kernel, allows one to resum leading (LL)
$\alpha_S^n\ln^n(\mu_F^2/\mu_{0F}^2)$ and next-to-leading (NLL) 
$\alpha_S^n \ln^{n-1}(\mu_F^2/\mu_{0F}^2)$ logarithms (collinear
resummation).
The explicit expression for the solution of the DGLAP equations can be found,
for instance, in Ref.~\cite{mele}.
Setting $\mu_{0F}\simeq m_b$ and $\mu_F\simeq Q$, one resums the large
$\ln(Q^2/m_b^2)$ appearing in the massive spectrum (\ref{massb}).

The NNLO initial condition of the perturbative fragmentation function
was calculated in Refs.~\cite{alex1,alex2}; 
however, for the NNLO result to be applicable, one would also
need three-loop time-like splitting functions, which are currently 
unknown.

Furthermore, both the initial condition (\ref{dbb}) and the coefficient 
functions in \cite{mele,cm,cor} present terms, $\sim 1/(1-x_b)_+$ and
$\sim [\ln(1-x_b)/(1-x_b)]_+$, which become large for $x_b\to 1$.
The large-$x_b$ limit corresponds to soft-gluon radiation; in Mellin
moment space, such terms correspond, at ${\cal O}(\alpha_S)$, to
single ($\sim\alpha_S\ln N$) and double ($\sim\alpha_S\ln^2 N$) logarithms of
the Mellin variable $N$.
Soft resummation in the initial condition of the perturbative fragmentation
function is process-independent and was performed in \cite{cc} in the
NLL approximation. Large-$x_b$ resummation in the coefficient
functions was implemented in \cite{cc} for $e^+e^-$ annihilation,
in \cite{ccm} for top decay, in \cite{cor} for $H\to b\bar b$ processes.
In $N$-space and in the
NLL approximation, terms $\sim \alpha_S^n\ln^{n+1}N$ (LL) and
$\sim \alpha_S^n\ln^nN$ (NLL) are kept in the Sudakov exponent.

The soft and collinear resummations in 
\cite{cc,ccm,cor} are matched to the exact NLO results, in order to 
give a reliable prediction over the full $x_b$ ($N$) range. This implies
 that the total widths are NLO as well.

It was shown in Refs.~\cite{cc,ccm,cor} that, after soft and collinear
logarithms are resummed, the $b$-quark spectra present very little
dependence on factorization and renormalization scales, which is 
equivalent to saying that the theoretical uncertainty is reduced.
Hereafter, we shall set the scales appearing in Eqs.~(\ref{pff}) and
(\ref{dbb})
to: $\mu=\mu_F=Q$, $\mu_0=\mu_{0F}=m_b$.

\section{Parton shower algorithms}
In this section we discuss parton shower algorithms, as implemented in
Monte Carlo event generators, such as HERWIG and PYTHIA.
These algorithms rely on the universality of the elementary branching
probability in the soft or collinear approximation.
Referring, for simplicity, to final-state radiation,
the probability of emission of a soft or collinear parton 
reads:
\begin{equation}
  \label{elem}
  dP={{\alpha_S}\over{2\pi}}{{dQ^2}\over{Q^2}}\ 
  \hat P(z)\  dz\ 
 {{ \Delta_S(Q^2_{\mathrm{max}},Q^2)}\over {\Delta_S(Q^2,Q_0^2)}}.
\end{equation}
In (\ref{elem}), $\hat P(z)$ is the tree-level Altarelli--Parisi splitting
function \footnote{Unlike Eq.~(\ref{pqq}), which includes virtual
corrections as well, in Eq.~(\ref{elem}) 
$\hat P(z)$ only accounts for real, small-angle parton radiation.
We have, 
for example, $\hat P_{qq}(z)=C_F(1+z^2)/(1-z)$, without the plus prescription
appearing in Eq.~(\ref{pqq}).},  
$z$ is the energy fraction of the emitted parton with respect to the
emitter, $Q^2$ is the ordering variable of the shower.
In HERWIG \cite{her}, $Q^2$ is an energy-weighted angle \footnote{
In the HERWIG showering frame, $Q^2\simeq E^2(1-\cos\theta)$, where $E$ is the
energy of the splitting parton and $\theta$ is the emission angle
\cite{marweb}.}, 
which corresponds
to angular ordering in the soft limit \cite{marweb}.
In PYTHIA \cite{pythia62}, $Q^2$ is the momentum squared of
the radiating parton, with an option to veto branchings that do not
fulfil the angular ordering prescription. 
Moreover, the latest version of it,
PYTHIA 6.3 \cite{pythia63}, 
offers as an alternative the possibility to order 
final-state showers according to the transverse momentum of the emitted
parton with respect to the emitter's direction, along the lines of
\cite{skands}.
For most of the results that we shall present, 
we shall use PYTHIA 6.220, with the option to
reject non-angular-ordered showers turned on, and HERWIG 6.506
\cite{her65}.

In (\ref{elem}) $\Delta_S(Q_1^2,Q_2^2)$ is the Sudakov form
factor, expressing the probability of evolution from $Q_1^2$ to $Q_2^2$ 
with no resolvable emission. In particular, the ratio of form factors
in (\ref{elem}) represents the probability that the considered emission
is the first, i.e. there is no emission
between $Q^2$ and $Q^2_{\mathrm{max}}$, where $Q^2_{\mathrm{max}}$ is set by
the hard-scattering process. $Q_0^2$ is instead the value of $Q^2$ at which
the shower evolution is terminated.
In diagrammatic terms, the Sudakov form factor
sums up all virtual and unresolved real emissions to all orders.

For multiparton radiation, 
iterating the branching probability (\ref{elem}) is equivalent to performing
the resummation of soft- and collinear-enhanced radiation.
As discussed, for example, in \cite{cmw} in the framework of the HERWIG
event generator, parton shower algorithms resum leading logarithms in the
Sudakov exponent, and include a class of subleading NLLs as well.

Standard Monte Carlo event generators yield the leading-order total
cross section for all implemented processes. 
The more recent `Monte Carlo
at next-to-leading order' (MC@NLO) \cite{mcnlo} program
implements instead both real
and virtual corrections to the hard-scattering process, in such
a way that predicted observables, including total cross sections,
are correct to NLO accuracy.

While resummed calculations typically do not implement the widths of the
decaying particles, whose masses are fixed quantities in the calculations,
Monte Carlo generators include the finite widths of
$Z$, $W$, Higgs bosons and top quarks. Some discussion on width effects
will be presented in section 6, when we shall investigate the decay of a Higgs 
boson with a mass of 500 GeV.

Moreover, for Higgs or top quark events, both HERWIG and PYTHIA
neglect interference 
effects between production and decay phases. In this approximation, the
differential decay width, normalized to the total width,
is equal to the normalized differential cross section,
independently of the production process:
\begin{equation}
{1\over\Gamma} {{d\Gamma}\over{dx_b}}={1\over\sigma}{{d\sigma}\over{dx_b}}.
\label{inter}
\end{equation}
Neglecting interference is a reliable approximation, as long as
the energies of the radiated gluons are much larger than the top or Higgs
widths \cite{orr}. 
This is indeed the case 
whenever the experimental analyses set cuts, e.g.
on the transverse momenta of the reconstructed jets in top-quark events,
of the order of 10~GeV
or larger. Although not really essential in view of Eq.~(\ref{inter}),
in the following we shall nonetheless run HERWIG and PYTHIA
for top and Higgs production at the LHC, i.e. for $pp$ collisions
at a centre-of-mass energy $\sqrt{s}=14$~TeV.

\section{Matrix-element corrections to parton showers}
The algorithm briefly discussed in the previous section is able to
simulate soft or collinear radiation;  for hard and large-angle
emission, the exact
matrix element must be used. In fact, Monte Carlo parton showers are
supplemented by matrix-element corrections.

PYTHIA uses the soft/collinear approximation in all the physical
phase space and the exact tree-level 
matrix element corrects the first emission \cite{miu,norrbin}.
In particular, PYTHIA 6.220 
contains matrix-element corrections to all
processes that we shall investigate.

The standard HERWIG algorithm entirely suppresses radiation in the
so-called `dead zone' of the phase space, 
corresponding to hard and large-angle emission.
The exact matrix element is used to populate the dead zone (hard
correction) and to correct the
shower every time an emission is capable of being the `hardest so far'
(soft correction), i.e. 
the emitted parton has the largest transverse momentum with respect
to the emitting one \cite{mike}.
HERWIG 6.506 \cite{her65} contains matrix-element corrections to
$e^+e^-$ annihilation \cite{mike1}, Deep Inelastic Scattering
\cite{mike2}, top decay \cite{corsey1} and vector-boson production 
\cite{corsey2}.
The corrections to Higgs hadroproduction ($gg/q\bar q\to H$),
whose inclusion is in progress, are discussed in \cite{moretti}.
In this section, we shall briefly discuss the implementation of matrix-element
corrections to parton showers in $H\to b\bar b$ decays, not yet
present in HERWIG.

\begin{figure}[ht!]
\centerline{\resizebox{0.69\textwidth}{!}{\includegraphics{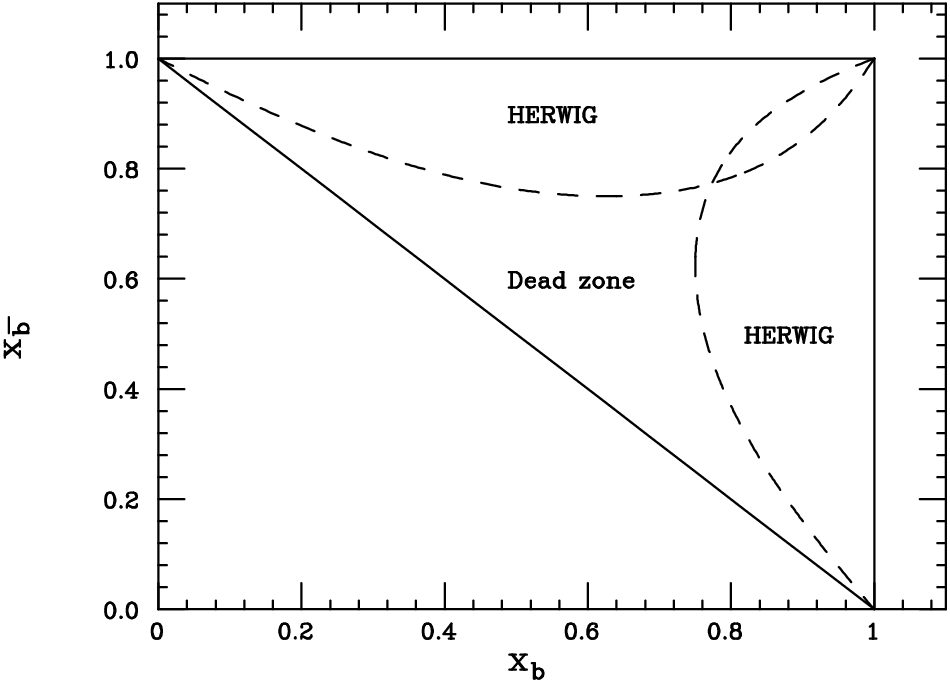}}}
\caption{Total (solid) and HERWIG (dashes) phase spaces for gluon radiation in
$H\to b\bar b$ processes. The dead zone is empty in the standard algorithm.}
\label{dead}
\end{figure}

As the Higgs is a colourless particle, the corrections to 
$H\to b\bar b$ are a straightforward extension of the ones to 
$Z\to b\bar b$ processes \cite{mike1}. The limits of the phase-space
region populated by HERWIG will be exactly the ones computed 
in \cite{mike1}, with $m_H$ replacing $m_Z$. 
The total and HERWIG's phase spaces 
are plotted in Fig.~\ref{dead}, in terms of $x_b$ and
$x_{\bar b}$. Collinear, corresponding to either $x_b=1$
or $x_{\bar b}=1$, and soft emissions, i.e. the point $x_b=x_{\bar b}=1$,
are within the HERWIG phase space; no radiation is allowed in the dead zone. 
We also note an overlapping region, where radiation may
come from either $b$ or $\bar b$.

To implement matrix-element corrections, we need to compute
the double-differential width of $H\to b\bar b g$ processes, 
corresponding to the Feynman diagrams in Fig.~\ref{bbg},
in terms of the $b$ and $\bar b$ energy fractions. It reads:
\begin{equation}
{1\over{\Gamma_0}}{{d^2\Gamma}\over{dx_b\ dx_{\bar b}}}=
{{\alpha_SC_F}\over{2\pi}}\ 
{{x_b^2+x_{\bar b}^2+2x_bx_{\bar b}-
2x_b-2x_{\bar b}+2}\over{(1-x_b)(1-x_{\bar b})}}.
\label{xbbar}
\end{equation}
In Eq.~(\ref{xbbar}) the scale of $\alpha_S$ is set to the gluon
transverse momentum with respect to the direction of the emitting quark:
this choice is suitable to
sum up a class of next-to-leading logarithmic corrections
\cite{ct}.
When applying matrix-element corrections,
we generate events in the dead zone according to Eq.~(\ref{xbbar}), and
use the exact result to correct the cascade in the
already-populated region,
any time an emission is the
hardest so far. We have estimated that, after matrix-element corrections  
are applied, for a Higgs mass $m_H=120$~GeV, about $4\%$ of events have
a gluon emission in the dead zone. For $m_H=500$~GeV, this fraction becomes
about $3\%$, since, for a larger mass value, even the gluon
transverse momentum $q_T^2$ is on average 
larger and therefore $\alpha_S(q_T^2)$ smaller.

\vspace{1.cm}\input feynman
\begin{figure}[ht!]
\begin{center}
\begin{picture}(24000,4000)
\THICKLINES
\drawline\scalar[\E\REG](-3000,0)[2]
\drawarrow[\LDIR\ATTIP](\pmidx,\pmidy)
\global\advance\pfronty by 500
\put(\pfrontx,\pfronty){$H$\ }
\drawline\fermion[\NE\REG](\pbackx,\pbacky)[6000]
\put(\pbackx,\pbacky){\ \ $b$}
\drawarrow[\LDIR\ATTIP](\pmidx,\pmidy)
\drawline\fermion[\SE\REG](\pfrontx,\pfronty)[3000]
\drawarrow[\NW\ATTIP](\pmidx,\pmidy)
\drawline\gluon[\NE\REG](\pbackx,\pbacky)[2]
\put(\pbackx,\pbacky){\ \ $g$}
\drawline\fermion[\SE\REG](\pfrontx,\pfronty)[3000]
\drawarrow[\NW\ATTIP](\pmidx,\pmidy)
\put(\pbackx,\pbacky){\ \ $\bar b$}
\drawline\scalar[\E\REG](18000,0)[2]
\drawarrow[\LDIR\ATTIP](\pmidx,\pmidy)
\global\advance\pfronty by 500
\put(\pfrontx,\pfronty){$H$\ }
\drawline\fermion[\SE\REG](\pbackx,\pbacky)[6000]
\drawarrow[\NW\ATTIP](\pmidx,\pmidy)
\put(\pbackx,\pbacky){\ \ $\bar b$}
\drawline\fermion[\NE\REG](\pfrontx,\pfronty)[3000]
\drawarrow[\LDIR\ATTIP](\pmidx,\pmidy)
\drawline\gluon[\SE\REG](\pbackx,\pbacky)[2]
\put(\pbackx,\pbacky){\ \ $g$}
\drawline\fermion[\NE\REG](\pfrontx,\pfronty)[3000]
\drawarrow[\LDIR\ATTIP](\pmidx,\pmidy)
\put(\pbackx,\pbacky){\ \ $b$}
\end{picture}\vspace{2.cm}
\caption{Feynman diagrams for $H\to b\bar b g$.}
\label{bbg}
\end{center}
\end{figure}
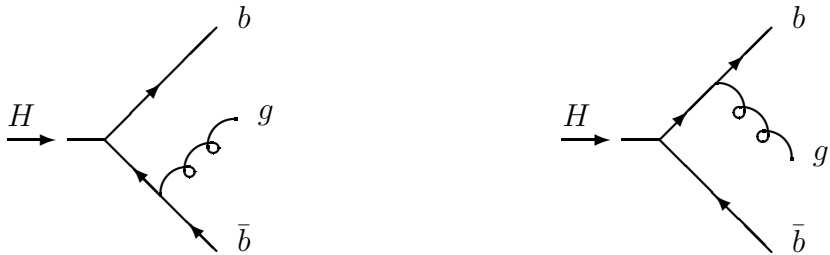
\begin{figure}[ht!]
\centerline{\resizebox{0.49\textwidth}{!}{\includegraphics{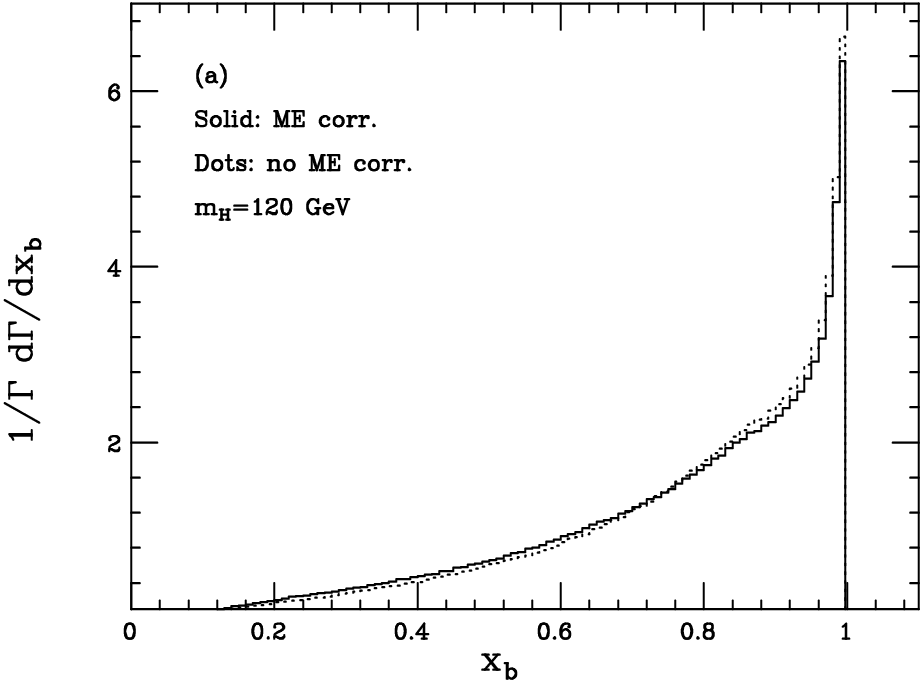}}%
\hfill%
\resizebox{0.49\textwidth}{!}{\includegraphics{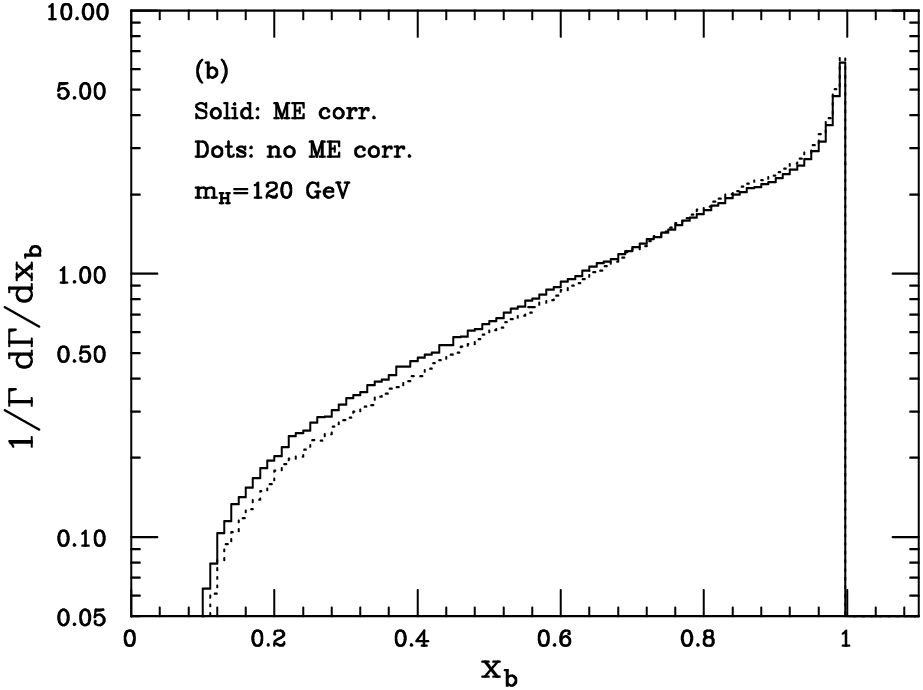}}}
\caption{$b$-quark energy spectrum in Higgs decay, according to
HERWIG with (solid) and without (dotted) matrix-element corrections,
on linear (a) and logarithmic (b) scales, for
$m_H=120$~GeV.}
\label{mecx}
\end{figure}
\begin{figure}[ht!]
\centerline{\resizebox{0.49\textwidth}{!}{\includegraphics{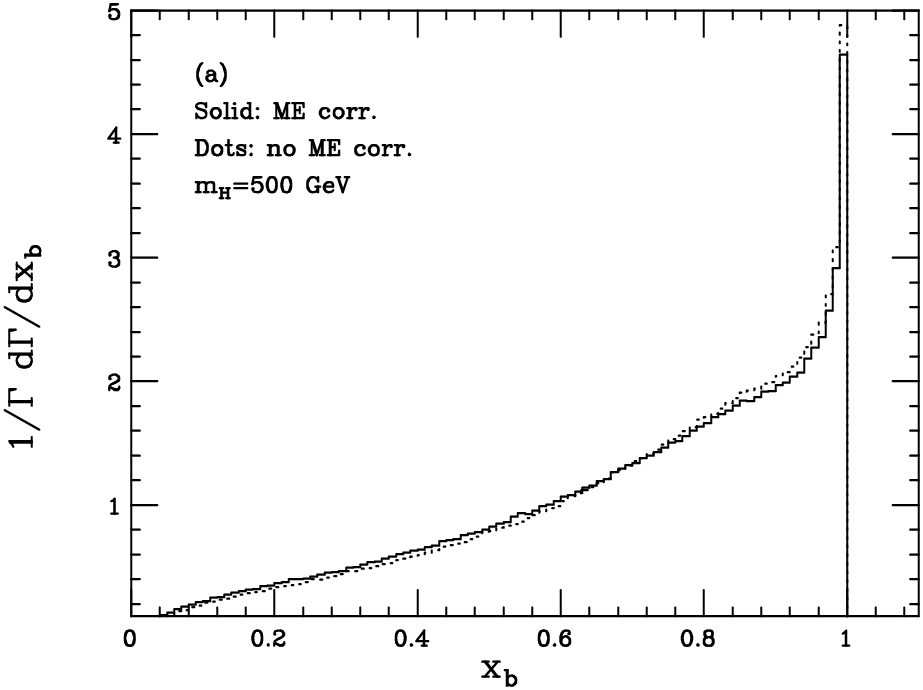}}%
\hfill%
\resizebox{0.49\textwidth}{!}{\includegraphics{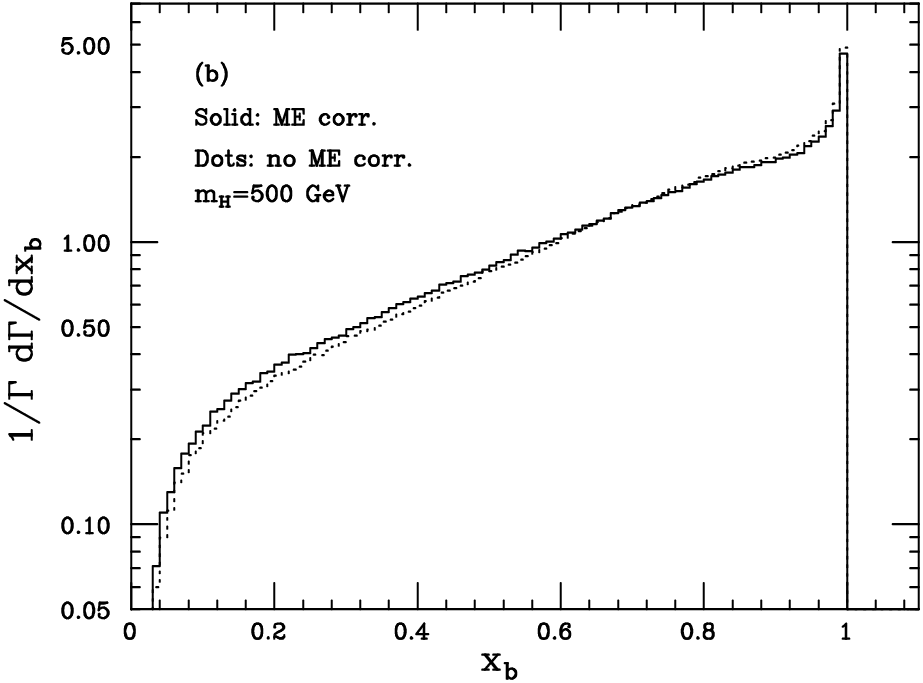}}}
\caption{As in Fig.~\ref{mecx}, but for a Higgs mass $m_H=500$~GeV.}
\label{mec500}
\end{figure}
\par
Figures~\ref{mecx}--\ref{mec500} 
show the effect of matrix-element corrections on the
$b$-quark energy fraction, for $m_H=120$ and 500 GeV, the values that
we will consider in the following.
As expected, more events are generated at
small $x_b$ (corresponding to the dead zone of the standard algorithm)
through the exact amplitude; this enhancement is compensated by 
having less events at large $x_b$.
In fact,
even after matrix-element corrections are applied, HERWIG, as well
as PYTHIA, 
gives by default the total leading-order cross sections (widths).
We also observe that the $x_b$ spectra present a sharp peak for $x_b\to 1$, 
which corresponds to events where no gluon is emitted by the $b\bar b$
pair ($x_b=x_{\bar b}=1$).
The fraction of events with $x_b=1$ depends on the shower 
cutoff $Q_0$ in Eq.~(\ref{elem}),
a user-defined parameter of HERWIG \cite{her}.

\section{$B$-hadron spectrum in $x_B$-space}
In this section we wish to present results on the $B$-hadron spectrum
in $e^+e^-$ annihilation, Higgs and top decay. 
Our tools to describe $b$-quark production will be, as discussed above, 
NLL-resummed calculations, HERWIG and PYTHIA.
Given the different perturbative accuracies,
treatment of width effects and hadronization models
which are implemented,
we do not expect that NLL computations and Monte Carlo event generators
should necessarily agree. However, we find it very interesting to perform
this comparison, since they are tools that are used for the analyses
of $b$-quark fragmentation, and it is useful to
estimate the quality of the agreement between the predictions they make.
Since the description of perturbative $b$ production is different,
in order for our comparison to be consistent, we shall have to
tune independently HERWIG, PYTHIA and the hadronization model used in the
framework of resummed calculations to the same data set.
Such a consistency is crucial if we wish to extract non-perturbative
information from $e^+e^-$ annihilation and use it to predict
$B$-hadron production in other processes \cite{canas,mlm}.

In order to describe $B$-hadron spectra in the considered processes,
we use experimental data on $B$ production in $e^+e^-$ annihilation, and rely
on the universality of the hadronization mechanism.
We shall account for data from SLD \cite{sld} and from 
the LEP experiments ALEPH \cite{aleph} and OPAL \cite{opal}
\footnote{Contrary to Refs.~\cite{cm,ccm,cor}, 
here we consider the OPAL data too.}.
Our results will be expressed in terms of the $B$ normalized energy fraction 
$x_B$, which is the hadron-level counterpart of Eq.~(\ref{xb}):
\begin{equation}
x_B={1\over{1-w}}{{2p_B\cdot Q}\over{Q^2}},
\label{xbhad}
\end{equation}
where $p_B$ is the $B$ four-momentum. As pointed out in section 2,
when we use a resummed calculation that neglects powers of $m_b^2/Q^2$,
$w=m_W^2/m_t^2$ for top decay, while $w=0$ for the other processes.
Since Monte Carlo event generators include some $m_b^2/Q^2$ effects, 
$w=m_W^2/m_t^2-m_b^2/m_t^2$ for top decays in HERWIG and PYTHIA.

ALEPH and OPAL reconstructed only weakly-decaying
$B$ mesons, while the SLD sample 
contains some $B$ baryons too; however, since the fraction of baryons
is pretty small, we can safely fit all data together and, when running
HERWIG or PYTHIA, we shall account for both mesons and baryons. 
As the resummed calculation yields
a NLO total cross section (width), and the considered
Monte Carlo programs only the LO one,
we shall study normalized distributions, such as 
$1/\Gamma\  d\Gamma/dx_B$, everywhere.

As far as the NLL resummed calculation is concerned, up to power 
corrections, one writes the hadron-level spectrum as 
the convolution of the parton-level one, determined as discussed in Section 2,
with a non-perturbative fragmentation function:
\begin{equation}
{1\over {\Gamma}} {{d\Gamma_B}\over{dx_B}} (x_B,Q,m_b)={1\over{\Gamma}}
\int_{x_B}^1 {{{dz}\over z}{{d\Gamma_b}\over {dz}}(z,Q,m_b)
D^{\mathrm{np}}\left({x_B\over z}\right)}.
\label{npff}
\end{equation}
We describe the hadronization using
the Kartvelishvili non-perturbative fragmentation function
\cite{kart}:
\begin{equation}
D^{\mathrm{np}}(x;\gamma)=(1+\gamma)(2+\gamma) (1-x) x^\gamma,
\label{kk}
\end{equation}
and fit the free parameter $\gamma$ to the data. 
As in \cite{cm,ccm,cor}, we consider the data in the range
$0.18\leq x_B\leq 0.94$, to avoid the regions $x_B\to 0$ and $x_B\to 1$, where
the calculation is unreliable. In fact, the parton- and hadron-level spectra
presented in \cite{cc,cm,ccm,cor} become negative at very small and very 
large $x$, which is due to the presence of terms that
have not been resummed yet, and to non-perturbative corrections,
relevant especially at large $x$. When doing the fit, we set the scales 
in Eqs.~(\ref{pff}) and (\ref{dbb}) and
the parameters entering the perturbative calculation to the following values:
$\mu=\mu_F=m_Z$, $\mu_0=\mu_{0F}=m_b$,
$m_Z=91.188$~GeV, $m_b=5$~GeV, $\Lambda^{(5)}_{\overline{\mathrm{MS}}}
=200$~GeV.
In the considered range, we obtain:
$\gamma=17.178\pm 0.303$,
with $\chi^2/\mathrm{dof}=46.2/53$ from the fit.

If we try to compare the data with the considered Monte Carlo
generators, we find that
the default parametrizations of the hadronization models in
HERWIG and PYTHIA are not able to reproduce the 
$x_B$ data. The $\chi^2$ per degree of freedom are, in fact, 
$\chi^2/\mathrm{dof}=739.4/61$ for HERWIG and $\chi^2/\mathrm{dof}=467.9/61$ 
for PYTHIA. A fit of HERWIG to a number of observables in
$e^+e^-$ annihilation was performed in \cite{hem}. 
Employing the parameters suggested in \cite{hem} gives
$\chi^2/\mathrm{dof}=391.9/61$ for the comparison with the $x_B$ data.

In this paper we wish to reconsider the tuning of HERWIG and PYTHIA to the
$b$-fragmentation data.
We point out that we are aware of the fact
that care must be taken when changing the
parameters of a Monte Carlo event generator,
since this may spoil the agreement with the data sets that
were used in the default tuning. The fits that we shall perform
should not be seen as an attempt
to redefine the parameters of HERWIG or PYTHIA; we would just like 
to investigate whether it is possible to change few parameters to describe
the $x_B$ data better. Of course, it will be interesting to use the
parametrization which we shall suggest to check how HERWIG and PYTHIA
fare with respect to other observables.
When doing the tuning, we 
concentrate on the parameters that determine the hadronization
and leave unchanged the ones related to the perturbative phase
of the showers and the quark masses.

In HERWIG, we modify the values of CLSMR(1) and CLSMR(2), controlling
the Gaussian smearing of the hadron direction with respect to
the original constituent quarks; PLSPLT(2), which determines the
mass distribution of $b$-flavoured cluster decays;
DECWT, affecting the
relative weight of decuplet and octet baryons; and CLPOW, to which
the heavy-cluster yield and the baryon/meson ratio are sensitive.
The fitted values are: CLSMR(1) = 0.4 (default 0), CLSMR(2) = 0.3 (0),
PSPLT(2) = 0.33 (1), DECWT = 0.7 (1), CLPOW = 2.1 (2). The values
of CLSMR(1), PSPLT(2) and DECWT are indeed the same as in Ref.~\cite{hem}.
After the tuning, the agreement with the data is still not very good,
but it is much better than with the default parametrization.
We find $\chi^2/\mathrm{dof}=222.4/61$.

\begin{figure}[ht!]
\centerline{\resizebox{0.69\textwidth}{!}{\includegraphics{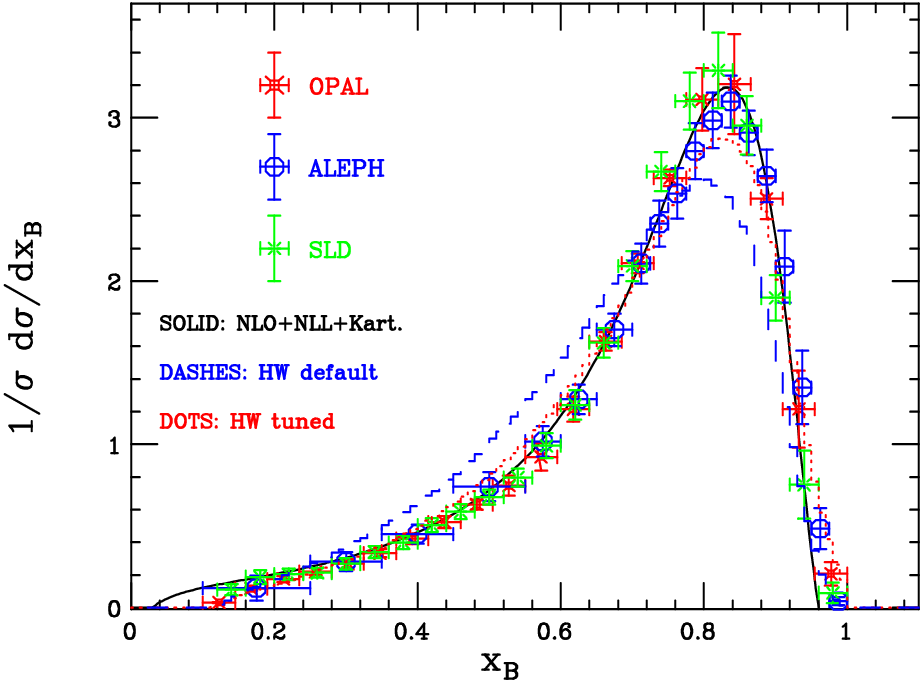}}}
\caption{Data from LEP and SLD experiments, compared with the NLO+NLL
calculation convoluted with the Kartvelishvili model (solid) 
and HERWIG 6.506, using the default parametrization (dashed)
and our tuning (dotted).}
\label{eeh}
\end{figure}

\begin{figure}[ht!]
\centerline{\resizebox{0.69\textwidth}{!}{\includegraphics{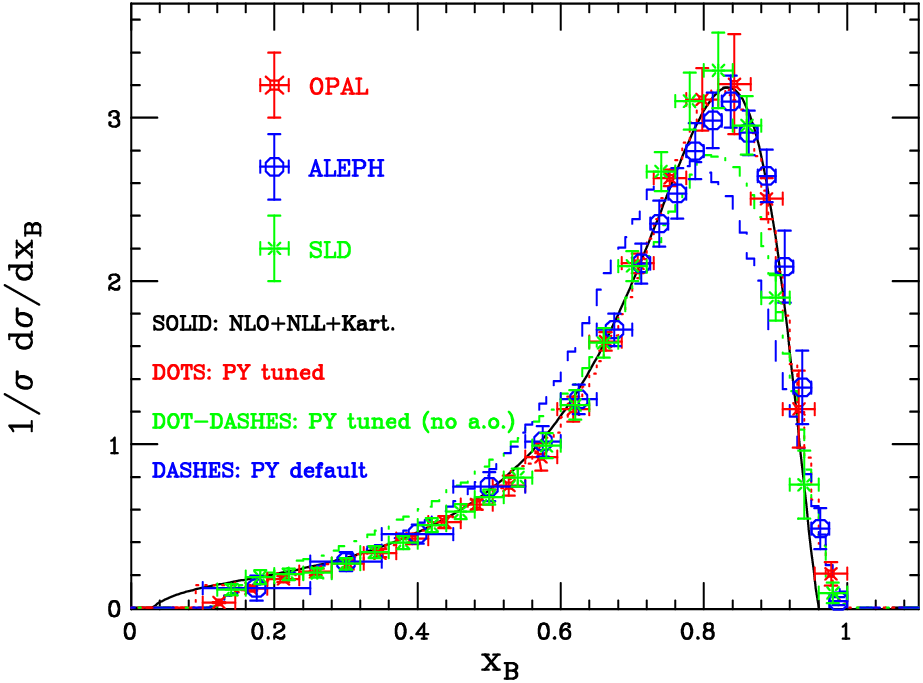}}}
\caption{As in Fig.~\ref{eeh}, but comparing data and the NLO+NLL
calculation with default (dashed) and tuned (dotted) PYTHIA 6.220.
Also shown is the PYTHIA prediction, using our tuning of the hadronization
model, but without rejecting the showers which do not
fulfil angular ordering (dot-dashed).}
\label{eep}
\end{figure}

As for PYTHIA, we change the values of the fragmentation
parameters
PARJ(41) and PARJ(42), which control the
$a$ and $b$ parameters of the Lund symmetric fragmentation function
\footnote{For a quark of energy $E$ moving on the $z$ axis, the quantity
$z$ appearing in Eq.~(\ref{lund}) is the fraction of $E+p_z$ taken
by the hadron, while 
$m_T$ is transverse mass, i.e. $m_T^2=m^2+p_x^2+p_y^2$.}:
\begin{equation} 
f_L(z)\propto {1\over z}(1-z)^a\exp(-bm_T^2/z).
\label{lund}
\end{equation}
We also tune PARJ(46), which modifies the endpoint of the Lund function 
according to the Bowler hadronization model \cite{bowler}
\footnote{See also the PYTHIA manual \cite{pythia62} for details on
the implementation of Eqs.~(\ref{lund}) and (\ref{bow}).}: 
\begin{equation}
f_B(z)\propto{1\over{z^{1+bm^2_q}}}(1-z)^a\exp(-bm_T^2/z),
\label{bow}
\end{equation}
where $m_q$ is the quark mass.
Our tuning gives the following values: 
PARJ(41) = 0.85 (default value 0.3), PARJ(42) = 1.03 (0.58), PARJ(46) = 
0.85 (1).
After our tuning, PYTHIA matches the $e^+e^-$ data well, 
and we obtain $\chi^2/\mathrm{dof}$ = 45.7/61 from the fit.
In Table~\ref{para} we summarize the parameters of
HERWIG and PYTHIA that we have changed in our tuning.
We have checked that our tuning works well also for the
new model
implemented in PYTHIA 6.3, as long as one runs it using 
options and parameters as they are defined in the new scenario (model 1) 
\cite{web}. 
Running PYTHIA 6.3, we find $\chi^2/\mathrm{dof}=46.0/61$
for the comparison with the $x_B$ data.
\begin{table}[t]
\caption{\label{para} Parameters of HERWIG and PYTHIA
hadronization models that we have tuned to improve the agreement
with $e^+e^-$ data, along with the $\chi^2$ per degree of freedom.}
\begin{center}
\begin{tabular}{|c|c|}\hline
HERWIG & PYTHIA \\
\hline\hline
CLSMR(1) = 0.4  &                 \\
 \hline
CLSMR(2) = 0.3  & PARJ(41) = 0.85 \\
\hline
DECWT = 0.7     & PARJ(42) = 1.03 \\
\hline
CLPOW = 2.1     & PARJ(46) = 0.85 \\
\hline
PSPLT(2) = 0.33 &                \\
\hline
\hline
$\chi^2/\mathrm{dof}$ = 222.4/61 & $\chi^2/\mathrm{dof}$ = 45.7/61 \\
\hline
\end{tabular}
\end{center}
\end{table}
In Figs.~\ref{eeh}-\ref{eep} we compare LEP and SLD data with HERWIG
and PYTHIA respectively, using the default and tuned parameters.
In both figures, we also show the results given by the resummed 
calculations of Ref.~\cite{cc}, convoluted with the 
Kartvelishvili model (solid), and denoted by `NLO+NLL+Kart.'
For the sake of comparison, we present the PYTHIA spectrum obtained
using the parameters of Table~\ref{para}, but without the rejection 
of non-angular-ordered showers. 

The $x_B$ 
distributions given by HERWIG and PYTHIA with the default parameters
deviate significantly from the data points, as
suggested by the $\chi^2$ values.
The NLO+NLL calculation and PYTHIA after the tuning
describe the data quite well. If we turn angular ordering off, PYTHIA
is obviously not capable of reproducing the data.
In fact, when tuning PYTHIA, we have used a version
that vetoes non-angular-ordered-showers;
if we allow non-angular-ordered branchings in PYTHIA, we should
in principle even reconsider the tuning. 
Besides the comparison with the data, it 
is nonetheless interesting to observe the effect of possible
non-angular-ordered showers: more events in the 
middle--low range $0.2\lsim x_B\lsim 0.6$ and less events around the peak.
In the following, we shall always run PYTHIA with forced angular ordering.

\par
The distribution yielded by HERWIG is instead 
somewhat broader than the data set. Even after the tuning,
HERWIG is above the data at small and large $x_B$ and below the data in the
neighbourhood of the peak. In any case, the comparison is greatly
improved with respect to the default parametrization, and in principle even 
the HERWIG parameters which control perturbative
$b$ production can be further tuned. Furthermore, major improvements 
in the treatment of $b$ fragmentation are present in HERWIG++ \cite{herplus},
the new object-oriented C++ version,
which uses new showering variables and includes quark masses in the splitting 
functions \cite{gies}. As discussed in \cite{herplus}, one can obtain
a quite good description of the SLD $x_B$ data fitting only the shower
cutoff, corresponding to $Q_0$ in Eq.~(\ref{elem}),
and without any additional tuning of the hadronization
model. The use of HERWIG++, whose first version simulates 
$e^+e^-$ annihilation, is however beyond the scope of this paper.

Relying on the universality of the hadronization mechanism, we can use the
parametrizations quoted in Table~\ref{para} to predict the
$x_B$ distribution in $H\to b\bar b$ and $t\to bW$. 
In Higgs decay we shall investigate the cases $m_H=120$ and 500 GeV;
in top decay we shall assume $m_t=175$~GeV and $m_W=80.425$~GeV.
The other quantities will be set consistently with the values
employed in the fit to the $e^+e^-$ data.

\begin{figure}[ht!]
\centerline{\resizebox{0.69\textwidth}{!}{\includegraphics{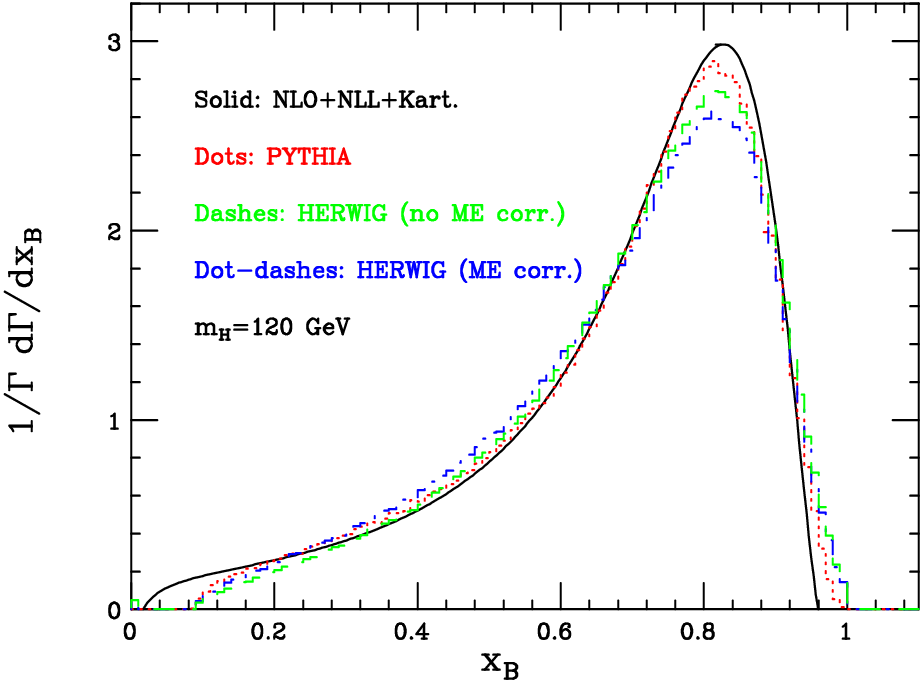}}}
\caption{$B$-hadron spectra in $H\to b\bar b$ processes,
according to a NLO+NLL calculation with the Kartvelishvili
non-perturbative model
(solid line), PYTHIA (dotted) and HERWIG with (dot-dashed) and without
(dashed) matrix-element corrections. We have set
$m_H=120$~GeV and used the tuning in Table~\ref{para}.}
\label{hph}
\end{figure}

In Fig.~\ref{hph}, we plot the $B$-hadron spectrum in
$H\to b\bar b$ processes, using HERWIG,
PYTHIA, and the NLL-resummed calculation
in \cite{cor}, convoluted with the Kartvelishvili model fitted as above.
We have set the Higgs mass to $m_H=120$~GeV.
PYTHIA fares rather well with respect the NLO+NLL calculation: 
although small discrepancies are present at very small and large $x_B$
and around the peak, the overall agreement looks acceptable.
The behaviour of the curves given by HERWIG reflects what we found in the
comparison with the $e^+e^-$ data: even in Higgs decay, the $x_B$ spectrum
is broader, lies above the NLO+NLL computation at large and
intermediate $x_B$, and below around the peak.
Since matrix-element corrections to $e^+e^-\to b\bar b$ were included when
we did the fit, in principle the tuned parameters must be used only when
matrix-element corrections to $H\to b\bar b$ are turned on.
However, for the sake of comparison, we show in Fig.~\ref{hph}
the $x_B$ spectrum given by HERWIG without such corrections as well.
The effect of matrix-element corrections is indeed what one should expect:
we have more events at small $x_B$, say $x_B<0.6$, and less at large $x_B$.
In fact, as shown in Figs.~\ref{mecx} and \ref{mec500},
the implementation of hard and large-angle gluon radiation 
enhances the probability to have 
$b$ quarks with smaller $x_b$, since the gluon is allowed to take
a larger energy fraction. It is therefore reasonable that this is reflected
by having more hadrons at small $x_B$ as well.

\begin{figure}[ht!]
\centerline{\resizebox{0.69\textwidth}{!}{\includegraphics{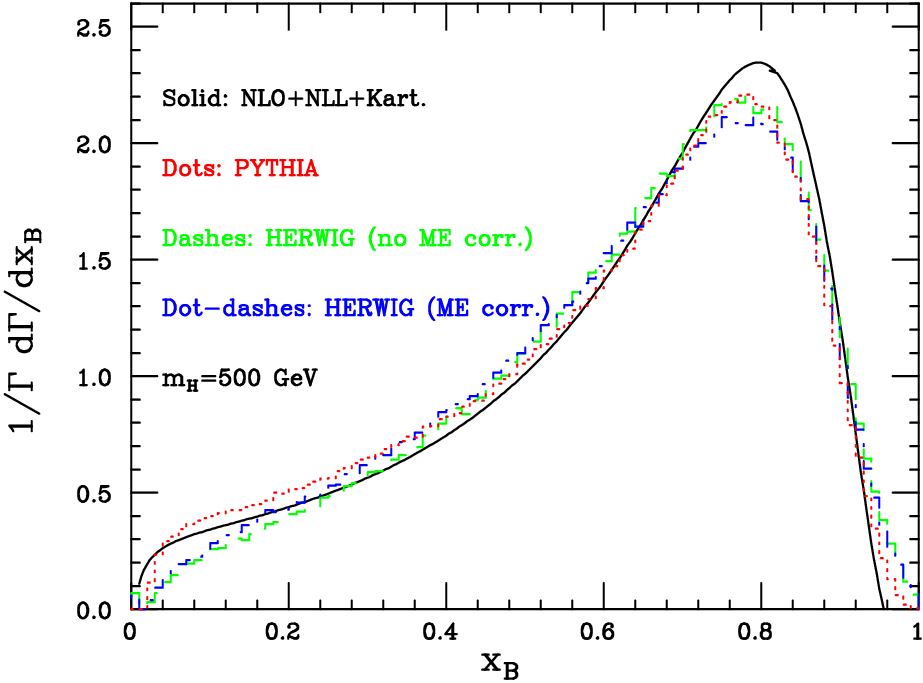}}}
\caption{As in Fig.~\ref{hph}, but for a Higgs mass $m_H=500$~GeV,
and with no width effects in HERWIG and PYTHIA.}
\label{hph500}
\end{figure}

We would like to consider also the case of a Higgs with a very large mass,
in particular $m_H=500$~GeV. This example does not have a
relevant application in 
the Standard Model, where the branching ratio of $H\to b\bar b$ is tiny for 
such a high-mass value,
but it could become more important in models with an extended Higgs 
sector. Another interesting aspect is to investigate whether 
our predictions are still 
in reasonable agreement for masses well above the $Z$-boson mass.
For $m_H=500$~GeV, width effects may become important; however, for
the sake of comparison with the resummed calculation,
which does not implement the Higgs width, we run HERWIG and PYTHIA for
$\Gamma_H=0$.
The spectra shown in Fig.~\ref{hph500} look broader than the 120 GeV case, 
but the comparison is rather similar: PYTHIA yields the highest
spectrum at small $x_B$ and HERWIG at intermediate and
very large $x_B$. The effects
of matrix-element corrections to $H\to b\bar b$ in HERWIG has an impact
that is similar to the one already observed if Fig.~\ref{hph}:
more events at small $x_B$ and less around the peak. We have 
finally checked that the
impact of the inclusion of finite-width effects in HERWIG and PYTHIA
is of the order of $1\%$. 
\par
\begin{figure}[ht!]
\centerline{\resizebox{0.69\textwidth}{!}{\includegraphics{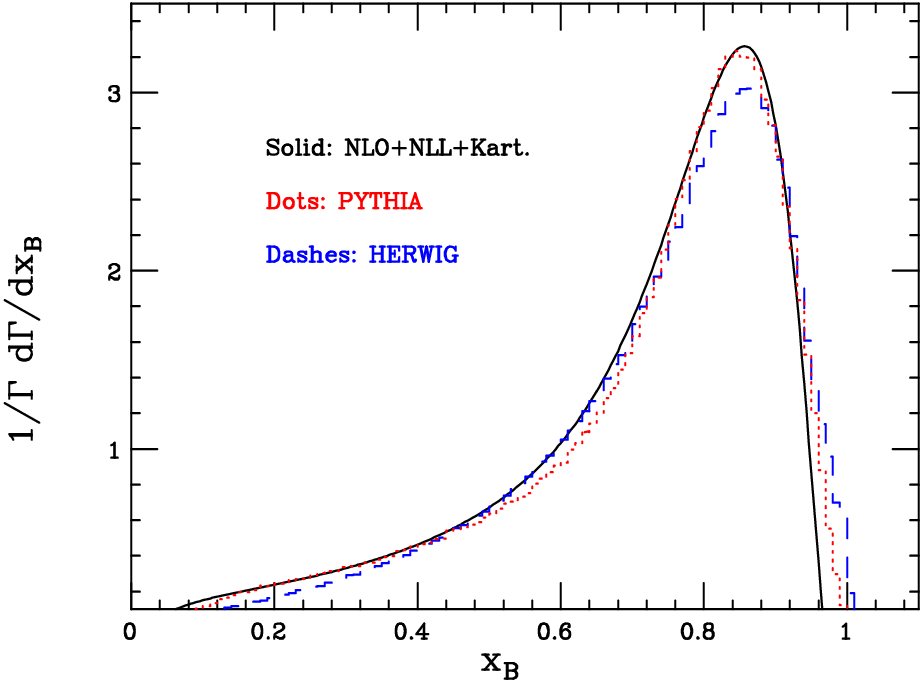}}}
\caption{$B$-hadron spectra in top decay, for $m_t=175$~GeV,
according to a NLO+NLL computation convoluted with the Kartvelishvili model,
(solid line), HERWIG (dashed) and PYTHIA (dotted).}
\label{hertop}
\end{figure}

In Fig.~\ref{hertop}, we compare the $B$-hadron spectrum in top-quark decay,
using HERWIG, PYTHIA  
and the perturbative NLO+NLL resummed calculation performed in
\cite{cm,ccm}, convoluted with the Kartvelishvili non-perturbative model.
PYTHIA is able to reproduce rather well the peak given by the
resummed calculation, while its spectrum
lies below it at $x_B\lsim 0.7$ and above 
it at $x_B\gsim 0.95$.
The spectrum given by HERWIG looks instead slightly different: it is 
below the resummed calculation in most of the $x_B$ range, i.e.
$x_B\lsim 0.9$, and above at very large $x_B$.
\begin{figure}[ht!]
\centerline{\resizebox{0.69\textwidth}{!}{\includegraphics{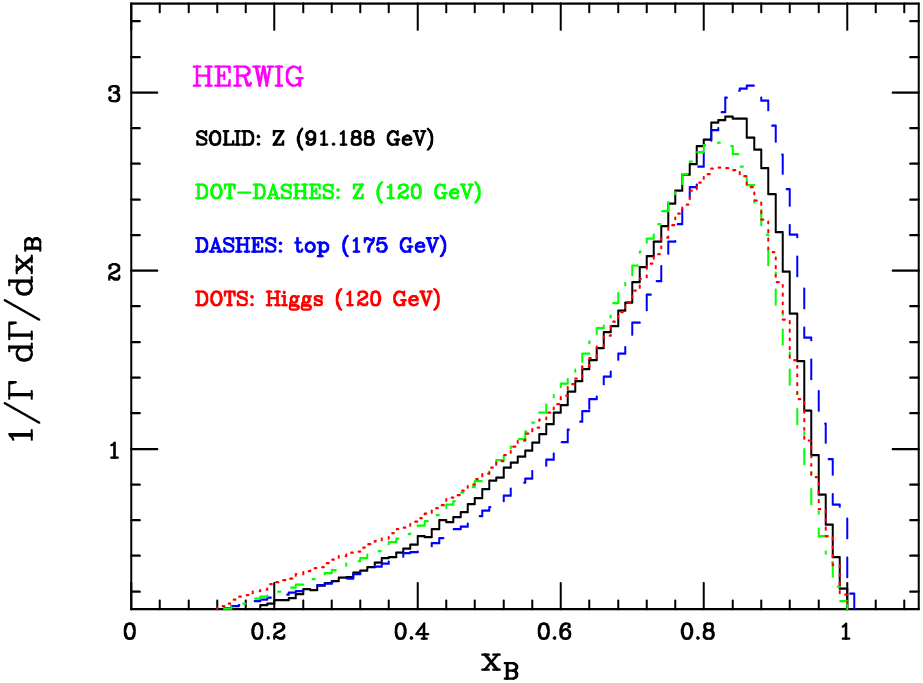}}}
\caption{Comparison of $B$-hadron spectra in $Z\to b\bar b$ for
$m_Z=91.188$~GeV (solid) and $m_Z=120$~GeV (dot-dashed),
for top decay (dashed) and for $H\to b\bar b$ (dotted), with $m_H=120$~GeV.  
Results given by HERWIG, with
matrix-element corrections included in all plots.}
\label{zbh}
\end{figure}
\begin{figure}[ht!]
\centerline{\resizebox{0.69\textwidth}{!}{\includegraphics{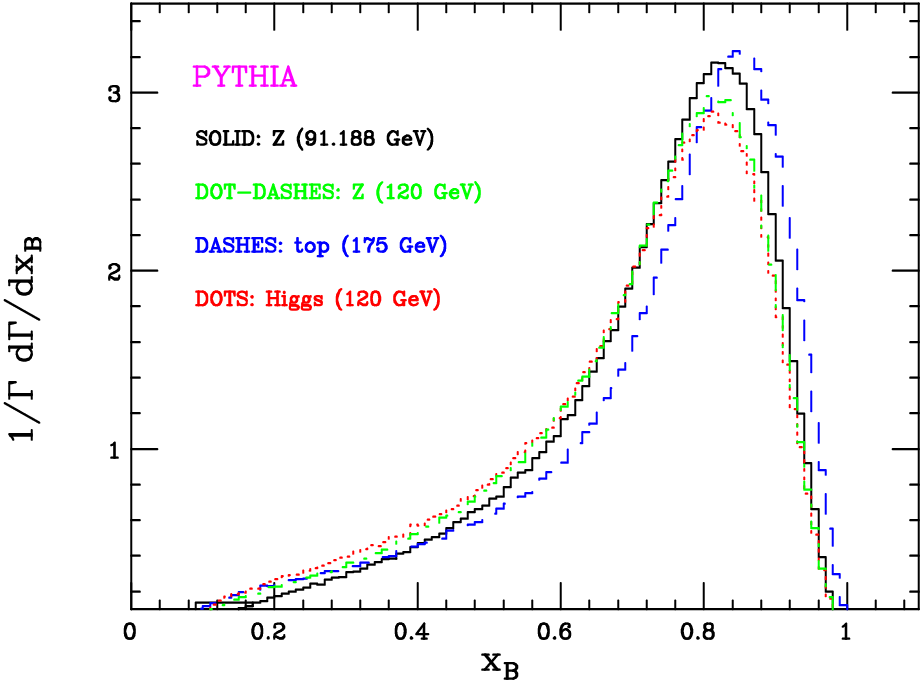}}}
\caption{As in Fig.~\ref{zbh}, but using PYTHIA.}
\label{zbhp}
\end{figure}
\par
In \cite{cor} the $B$ spectra in
$Z$, top and Higgs decays were compared
using NLL resummed calculations. In Figs.~\ref{zbh} and \ref{zbhp} 
we show the same 
comparison, but running HERWIG and PYTHIA; in order to investigate the effect
of the particle spins, we also show the result in $e^+e^-$ annihilation,
but for $\sqrt{s}=120$~GeV, the chosen value for the Higgs mass.
Figs.~\ref{zbh} and \ref{zbhp}
exhibit features similar to those of Ref.~\cite{cor}:
the $B$ energy distribution in top decay is shifted toward large $x_B$
values; in Higgs decay we have more events at small $x_B$; the spectrum
from $Z$ decay lies between the two. Setting $m_Z=m_H$, as observed in
\cite{cor}, makes the spectra from $H$ and $Z$ decays more similar, but
small discrepancies are still present.    
The comparison between top and Higgs decays is indeed quite relevant, since
the different $B$ spectra may help to distinguish between $b$-flavoured
hadrons in the process $pp\to t\bar t H$, followed by $H\to b\bar b$,
one of the most important channels for Higgs studies at the LHC
\cite{denegri}.

\section{Results in moment space}

In this section we
wish to present results in moment space, where the moments $\Gamma_N$ of a
function $1/\Gamma\ d\Gamma/dz$ are defined as follows: 
\begin{equation}
\Gamma_N  =\int_0^1 {dz  \ z^{N-1}
{1\over{\Gamma}}{{d\Gamma}\over{dz}}(z) }.
\label{moment}
\end{equation}
In the case of the histograms given by a Monte Carlo program, the integral
in (\ref{moment}) will be approximated to a discrete sum over the bins.
In Ref.~\cite{delphi}, the DELPHI collaboration presented the moments
for $B$ production in $e^+e^-$ annihilation.
From the point of view of resummed calculations, 
working in moment space \cite{canas,gardi} presents several advantages.
In $N$-space, convolutions become ordinary products, and the relation
between parton- and hadron-level cross sections becomes:
\begin{equation}
\sigma_N^B=\sigma_N^bD_N^{\mathrm{np}},
\end{equation} 
where $\sigma_N^b$ and $\sigma_N^B$ are the moments of the $b$ and $B$ cross 
sections, and $D_N^{\mathrm{np}}$ the $N$-space counterpart of the 
non-perturbative fragmentation function.
Therefore, there is no need to assume any functional form for the 
non-perturbative fragmentation function in $x$ space.
Moreover, resummed calculations 
are well defined in $N$-space, and we do not have the problem exhibited by the 
$x_B$-space spectra, which become negative at small or large $x_B$. 

In \cite{ccm,cor}, the parton- and hadron-level
moments for $t\to bW$ and $H\to b\bar b$ 
processes were presented, using a resummed NLL perturbative calculation.
In this paper, we wish to compare experimental data and predictions,
employing HERWIG and PYTHIA as well, tuned as in Table~\ref{para}.

In Table~\ref{tab} we quote the data from DELPHI, the moments yielded
by HERWIG and PYTHIA in $Z$, $t$ and $H$ decays, and, for the sake of 
comparison, the results, already published in \cite{ccm,cor}, yielded by
NLL-resummed calculations.
As for the NLL computation, we present two sets of results:
we multiply the moments of the perturbative cross sections (widths) by
the moments of the Kartvelishvili model, denoted by $K_N$ in 
Table~\ref{tab}, fitted to LEP and SLD
$x_B$-space data, and by
$D_N^{\mathrm{np}}$, extracted from the DELPHI $N$-space data \cite{delphi}.
\begin{table}[t]
\caption{\label{tab}\small  Moments
$\sigma^B_N$ from
DELPHI~\protect\cite{delphi}, and moments
in $e^+e^-$ annihilation, Higgs ($H$) and top ($t$) decay, 
using NLL resummed calculations, HERWIG (HW) and PYTHIA (PY).
In $H\to b\bar b$ , we consider the Higgs mass values
$m_H=120$~GeV and 500 GeV, and 
quote HERWIG results with (c.) and without (no c.)
matrix-element corrections. The NLL results are obtained
by extracting $D_n^{\mathrm{np}}$ from the DELPHI data,
and using the moments $K_N$ of the Kartvelishvili model, fitted 
to $x_B$ data, as in Figures \ref{eeh} and \ref{eep}.}
\small
\begin{tabular}{| c | c c c c |}
\hline
& $\langle x\rangle$ & $\langle x^2\rangle$ & $\langle x^3\rangle$
& $\langle x^4\rangle$ \\
\hline
\hline
$e^+e^-$ data $\sigma_N^B$&0.7153$\pm$0.0052 &0.5401$\pm$0.0064 &
0.4236$\pm$0.0065 &0.3406$\pm$0.0064  \\
\hline
\hline
$e^+e^-$ NLL $\sigma_N^b$   & 0.7801 & 0.6436 & 0.5479 & 0.4755  \\
\hline
$D^{\mathrm{np}}_N$ & 0.9169 & 0.8392 & 0.7731 & 0.7163 \\
\hline
$e^+e^-$ $\sigma_N^B=\sigma_N^bK_N$ 
   & 0.7027 & 0.5251 & 0.4066 & 0.3225  \\
\hline
$e^+e^-$ HW $\sigma_N^B$   & 0.7113 & 0.5354 & 0.4181 & 0.3353  \\
\hline
$e^+e^-$ PY $\sigma_N^B$   & 0.7162 & 0.5412 & 0.4237 & 0.3400  \\
\hline
\hline
$H$-dec.(120) NLL  $\Gamma^b_N$ & 0.7580 & 0.6166 & 0.5197 & 0.4477  \\
\hline
$H$-dec.(120) $\Gamma^B_N=\Gamma^b_N D_N^{\mathrm{np}} $
& 0.6950 & 0.5175 & 0.4018 & 0.3207 \\
\hline
$H$-dec.(120) $\Gamma^B_N=\Gamma_N^bK_N$
& 0.6829 & 0.5030 & 0.3858 & 0.3036 \\
\hline
$H$-dec.(120) HW c. $\Gamma^B_N$ & 0.6842 & 0.5036 & 0.3877 & 0.3076 \\
\hline
$H$-dec.(120) HW no c. $\Gamma^B_N$ & 
0.6961 & 0.5177 & 0.4011 & 0.3197 \\
\hline
$H$-dec.(120) PY $\Gamma^B_N$ & 0.6876 & 0.5080 & 0.3913 & 0.3099 \\
\hline
\hline
$H$-dec.(500) NLL  $\Gamma^b_N$ & 0.6858 & 0.5265 & 0.4255 & 0.3545  \\
\hline
$H$-dec.(500) $\Gamma^B_N=\Gamma^b_N D_N^{\mathrm{np}}$ 
& 0.6288 & 0.4418 & 0.3290 & 0.2539 \\
\hline
$H$-dec.(500)  $\Gamma^B_N=\Gamma_N^bK_N$ 
& 0.6178 & 0.4295 & 0.3158 & 0.2404 \\
\hline
$H$-dec.(500) HW c. $\Gamma^B_N$ & 0.6184 & 0.4279 & 0.3146 & 0.2406 \\
\hline
$H$-dec.(500) HW no c. $\Gamma^B_N$ & 0.6286 & 0.4389 & 0.3245 & 0.2491 \\
\hline
$H$-dec.(500) PY $\Gamma^B_N$ & 0.6044 & 0.4152 & 0.3036 & 0.2307 \\
\hline
\hline
$t$-dec. NLL $\Gamma^b_N$ & 0.7883 & 0.6615 & 0.5735 & 0.5071 \\
\hline
$t$-dec. NLL $\Gamma^B_N=\Gamma^b_N
D_N^{\mathrm{np}}$ & 0.7228 & 0.5551 & 0.4434 & 0.3632 \\
\hline
$t$-dec. $\Gamma^B_N=\Gamma_N^bK_N$ & 0.7102 & 0.5397 & 0.4257 & 0.3439 \\
\hline
$t$-dec. HW $\Gamma^B_N$ & 0.7325 & 0.5703 & 0.4606 & 0.3814 \\
\hline
$t$-dec. PY $\Gamma^B_N$ & 0.7225 & 0.5588 & 0.4486 & 0.3688 \\
\hline
\end{tabular}
\end{table}
The moments yielded by HERWIG and PYTHIA in $e^+e^-$ annihilation
are consistent, within the error
ranges, with the ones measured by DELPHI. 
Although problems are present when fitting the $x_B$ data
from LEP and SLD, it is remarkable that HERWIG is
compatible with the DELPHI moments within one standard deviation.
Moreover, we observe that the moments given by the NLL calculation are rather
different according to whether the fit is made directly to the $N$-space 
DELPHI data, or by fitting in $x_B$ space and then calculating the 
moments. In fact, the $x_B$-space fit
was performed in the range $0.18\leq x_B\leq 0.94$, while the moments
are computed integrating over the full range $0\leq x_B\leq 1$.
Neglecting data at small and large $x_B$ in the fit may
cause an incorrect determination of the moments, which do depend on the tails
of the distributions \cite{canas,gardi}.

The results for top and Higgs decay
exhibit similar features to the $x_B$ spectra.
In top decay, PYTHIA is very close to the NLL calculation which uses
$D_N^{\mathrm{np}}$ extracted from the DELPHI data, while
HERWIG, whose predictions are shifted toward larger $x_B$,
gives larger moments. For $H\to b\bar b$ and $m_H=120$~GeV, PYTHIA and
HERWIG, after matrix-element corrections, give moments which are compatible
within $1\%$, and are closer to the NLL result which uses the
Kartvelishvili model rather than the moments $D_N^{\mathrm{np}}$.
For $m_H=500$~GeV, HERWIG looks
somewhat closer to the NLL result, while PYTHIA gives moments
which are a few per cent smaller.
Matrix-element corrections to 
$H\to b\bar b$ in HERWIG have the effect to reduce the values of the moments
of about 2--5$\%$ for both $m_H=120$ and 500 GeV,
which is understandable, since more
small-$x_B$ events are generated after the inclusions of such corrections.

\section{Conclusions}
We have studied bottom-quark fragmentation in $e^+e^-$ annihilation,
as well as in top and Higgs decay, 
using NLL-resummed calculations based on the
fragmentation function formalism and Monte Carlo event generators, such as 
HERWIG and PYTHIA. Monte Carlo parton showers simulate multiple radiation in 
the soft or collinear approximation, and need to be provided with 
matrix-element corrections to simulate hard and large-angle parton radiation.
While PYTHIA contained the corrections to all three considered
processes, we had to
implement matrix-element corrections to $H\to b\bar b$ in
HERWIG, since they are not yet present in HERWIG. At parton level,
the effect of such corrections is that there are more $b$ quarks 
simulated with a smaller energy fraction.

We have fitted HERWIG, PYTHIA and the Kartvelishvili hadronization 
model, the latter used in conjunction with the NLL-resummed calculation, 
to $e^+e^-$ data on the $B$-hadron energy fraction $x_B$, 
from ALEPH, OPAL and SLD.
We found that the Kartvelishvili and the PYTHIA string
model are able to 
fit the data quite well, while the HERWIG cluster model
is only marginally consistent.
The tuning of the Monte Carlo programs turned out to be essential,
since the default parametrizations fare rather badly against 
the measured $x_B$ spectrum. 

Relying on the process-independence of the hadronization,
we have predicted the $x_B$ spectrum in Higgs and top decays, using the
tuned models, and found results which reflect the quality of the fit
to the $e^+e^-$ data. PYTHIA and resummed calculations give consistent
results, while the HERWIG predictions are broader and compatible only in 
some $x_B$ ranges.
The effect of matrix-element corrections to $H\to b\bar b$ is reasonable,
since more events are generated via the exact amplitude
at small $x_B$, which corresponds, at parton level,
to hard and large-angle gluon radiation.
Comparing the spectra of $B$ hadrons in $Z$, $H$ and top decays has
given similar results to the ones already found in the framework of 
perturbative fragmentation functions: the $B$'s from top quarks are the
hardest, the ones from $H$ are the softest, and the ones from $Z$ 
lie within the two.

Finally, we have considered data on the moments of the $B$ cross section
measured at LEP by the DELPHI collaboration. We have extracted the moments
of the non-perturbative fragmentation function, and  compared the
$N$-space cross sections given by the NLL computation to those yielded by
HERWIG and PYTHIA.
We found that tuned HERWIG and PYTHIA are able to reproduce the DELPHI
moments, within the quoted error range, even though HERWIG had problems
with fitting the $e^+e^-$ $x_B$ data.
The $N$-space predictions in top decay reflect the behaviour of
the $x_B$-space results: PYTHIA is closer to the NLL calculation, and
HERWIG yields larger moments. In $H\to b\bar b$, all given moments
are pretty similar for $m_H=120$~GeV, while, for $m_H=500$~GeV, 
HERWIG is closer
to the NLL prediction and PYTHIA yields smaller results.
Matrix-element corrections to HERWIG
simulations of $H\to b\bar b$ decrease the moment values, which is
due to the smaller fraction of events at large $x_B$.
We have also pointed out that the moments based on an NLL $x_B$-space fit, 
making use of the Kartvelishvili fragmentation model in
a range which discards very large and very small values of $x_B$, are 
different from the ones obtained after 
fitting the moments experimentally measured by DELPHI directly.

In conclusion, we believe that our studies and tunings can be a useful starting
point to address $b$-quark fragmentation at the Tevatron and LHC,
especially for the analyses of top-quark properties,
such as its mass, and for the Higgs
studies in the $H\to b\bar b$ channel. 
It will also be interesting to use the parametrizations which we have proposed
to study other hadron-level observables not necessarily related to
$b$-quark fragmentation. Moreover, while we deliberately did not
aim at fitting any perturbative parameter entering in the Monte
Carlo simulation or in the resummed calculation, it will be nonetheless
cumbersome comparing parton-level observables and
investigating whether further tuning may improve the comparison
with the $e^+e^-$ data and affect the hadron-level predictions.

\section*{Acknowledgements}
We are grateful to M. Cacciari who provided us with the computer code to 
perform inverse Mellin transforms and fit the Kartvelishvili
model to $e^+e^-$ data.
We acknowledge M.H. Seymour and T. Sj\"ostrand for discussions on
HERWIG and PYTHIA event generators.
Work supported in part by the European Community's Human Potential Programme 
under contract HPRN-CT-2002-00326, [V.D.].

\end{document}